\documentclass[11pt,a4paper]{article}

\usepackage[dvips]{color}
\usepackage{graphicx}
\usepackage{amsmath}    

\def\Journal#1#2#3#4{{#1} {\bf #2}, #3 (#4)}

\def\NPB{{\em Nucl. Phys.} B}
\def\PLB{{\em Phys. Lett.}  B}
\def\PRL{\em Phys. Rev. Lett.}
 
\def\PRD{{\em Phys. Rev.} D}

\def\EPC{{\em Euro. Phys. J.} C}

\allowdisplaybreaks[4]

\begin{document}
\title{\vspace{-15mm}Lepton flavor violation in Higgs boson decays}

\author{Koji TSUMURA\\
\em Department of Physics, Osaka University Toyonaka, Osaka 560-0043, Japan}
\date{}
\maketitle
\vspace{-7mm}
\begin{abstract}
We discuss lepton flavor violation (LFV) associated with 
tau leptons in the framework of the two-Higgs-doublet model.
Current data for rare tau decays provide substantial upper limits    
on the LFV Yukawa couplings in the large $\tan\beta$ region 
where $\tan\beta$ is the ratio of vacuum expectation values 
of the two Higgs doublets.
We show that measuring the LFV Higgs boson decays 
$h \rightarrow \tau^\pm \mu^\mp$
at future colliders 
can be useful to further constrain the LFV couplings 
especially in the relatively small $\tan\beta$ region. 
\end{abstract}
\section{Lepton flavor violation in the two-Higgs-doublet model}
Lepton flavor violation (LFV) is one of 
the most obvious signature of physics beyond the Standard Model (SM).
It can naturally appear in extend Higgs sectors~\cite{Osaka,LFVHIGGS}. 
Some theories
(supersymmetric models, the little Higgs model, the Zee model, etc.) 
predict such extended Higgs sectors. 
Therefore the LFV is an important probe to find the signal 
of new physics.

The Higgs sector of the two-Higgs-doublet Model(THDM) is expressed as
\begin{align}
-\mathcal{L}_{\text{Higgs}} 
=&
m_{1}^{2} \left| \Phi_{1} \right|^{2}
+
m_{2}^{2} \left| \Phi_{2} \right|^{2}
-
\left( m_{3}^{2} \Phi_{1}^{\dagger} \Phi_{2} + {\rm H.c.} \right) 
\nonumber \\
&
+
\frac{\lambda_{1}}{2} \left|\Phi_{1}\right|^{4}
+
\frac{\lambda_{2}}{2} \left|\Phi_{2}\right|^{4} 
+
\lambda_{3} \left|\Phi_{1}\right|^{2} \left|\Phi_{2}\right|^{2}
+
\lambda_{4} \left|\Phi_{1}^{\dagger} \Phi_{2} \right|^{2}
+
\left\{
 \frac{\lambda_{5}}{2}
 \left( \Phi_{1}^{\dagger} \Phi_{2} \right)^{2} + {\rm H.c.}
\right\}, 
\label{eq:LHiggs}
\end{align}
where $\Phi_{1}$ and $\Phi_{2}$ are 
the scalar iso-doublets with hypercharge $1/2$.
We consider the model in which the discrete $Z_2$ symmetry is explicitly 
broken in the leptonic Yukawa interaction, 
and treat the CP conserving Higgs potential with the soft-breaking 
parameter $m_3^2$.
There are eight degrees of freedom in the two Higgs doublet fields.
Three of them are absorbed by the weak gauge bosons via the Higgs
mechanism, and remaining five are physical states.
After the diagonalization of the mass matrices, they correspond to 
two CP-even ($h$ and $H$), 
a CP-odd ($A$), and a pair of charged  ($H^{\pm}$) Higgs bosons.
We define such that $h$ is lighter than $H$.
The eight real parameters $m_{1-3}^{2}$ and $\lambda_{1-5}$ 
can be described by the same number of physical parameters; i.e.,
the vacuum expectation value $v$ $(\simeq 246$ GeV), 
the Higgs boson masses 
$m_{h}^{}, m_{H}^{}, m_{A}^{}$ and  $m_{H^{\pm}}^{}$,  
the mixing angle $\alpha$ between the CP-even Higgs bosons, 
the ratio $\tan \beta$ 
($ \equiv \langle \Phi^{0}_{2} \rangle / \langle \Phi^{0}_{1}
\rangle$) 
of the vacuum expectation values for two Higgs doublets,
and the soft-breaking scale $M$ $(\equiv \sqrt{m_{3}^{2}/\sin\beta
\cos\beta})$ for the discrete symmetry.
Parameters of the model are constrained from experimental results
(the $\rho$ parameter\cite{rho}, $b\to s\gamma$\cite{bsg}) 
and also from requirements of theoretical consistencies such 
as vacuum stability and perturbative unitarity~\cite{VS,PU}.

The tau lepton associated LFV couplings are parameterized 
in the mass eigenbasis of each field as~\cite{Osaka,LFVHIGGS,kot}
\begin{align}
-\mathcal{L}_{\tau{\rm LFV}}
=&
\frac{m_{\tau}}{v \cos^{2}\beta}
 \left(
 \kappa^{L}_{3i} \overline{\tau} {\rm P}_{L} \ell_{i}
 +
 \kappa^{R}_{i3} \overline{\tau} {\rm P}_{R} \ell_{i}
 \right) \cos\left(\alpha - \beta \right) h + {\rm H.c.},
\label{eq:tauLFV}
\end{align}
where ${\rm P}_{L}$ (${\rm P}_{R}$) is the projection operator to 
the left(right)-handed  field, and 
$\ell_1$ and $\ell_2$ represent $e$ and $\mu$ respectively.
When a new physics model is specified at the high energy scale, 
$\kappa^{L,R}_{ij}$ can be predicted as a function of the model 
parameters. 
\section{Rare tau decay results and LFV Higgs boson decay} 
We concentrate on the LFV coupling 
$|\kappa_{32}|$, where 
$|\kappa_{32}|^{2} \equiv |\kappa^{L}_{32}|^{2} +
|\kappa^{R}_{23}|^{2}$.
We take into account the data for rare tau decay processes 
such as 
$\tau \rightarrow \mu P^0 $,
$\tau \rightarrow \mu M^+M'^-$, 
$\tau \rightarrow \mu\ell^+\ell^-$, and   
$\tau \rightarrow \mu\gamma$,
where $P^0$ represents $\pi^0$, $\eta$ and $\eta'$ mesons, and 
$M^\pm$ ($M'^\pm$) does $\pi^\pm$ and $K^\pm$ mesons.
Since branching ratios  
depend on different combinations of the Higgs boson masses,
independent information can be obtained for the model parameters 
by measuring each of them. 
When all the masses of Higgs bosons are large,
these decay processes decouple by a factor of $1/m_{\rm Higgs}^{4}$.
These branching ratios are complicated functions 
of the mixing angles,
each of them can be simply expressed to be proportional 
to $\tan^{6} \beta$ for
$\tan\beta \gg 1$ in the SM like region ($\sin(\alpha-\beta) \sim -1$).
This $\tan^{6} \beta$ dependence is a common feature of the tau-associated LFV processes with the Higgs-mediated 4-Fermi interactions.  

The experimental upper limit 
on $|\kappa_{32}|^{2}$ can be obtained  
by using the experimental results and analytic
expressions for the decay branching ratios.
For instance, let us consider the bound 
from the $\tau\rightarrow \mu\eta$ results. 
We define the maximal allowed value for $|\kappa_{32}|^2$
\begin{align}
\left( 
 \left|
  \kappa^{\text{max}}_{32}
 \right|^{2}
 \right)_{\tau^{}\rightarrow \mu^{}\eta}
 & \hspace{-2.3mm} \equiv
 \frac{256 \pi {\rm Br}(\tau^{} \rightarrow \mu^{} \eta)_{\rm exp}  m_{A}^{4}}
 {9 G_{F}^{2} m_{\tau}^{3} m_{\eta}^{4} F_{\eta}^{2} \tau_{\tau} 
  \left(
   1- \frac{m_{\eta}^{2}}{m_{\tau}^{2}}
  \right)^{2}}
 \frac{\cos^{6}\beta}{\sin^{2}\beta}, 
\label{eq:bound-from-tmeta}
\end{align}
where  
$|\kappa_{32}|^{2} \leq
 \left( 
 \left|
  \kappa^{\text{max}}_{32}
 \right|^{2}
 \right)_{\tau^{}\rightarrow \mu^{}\eta}$
, $G_F$ is the Fermi constant, 
$F_\eta$ is the decay constant for $\eta$, 
and ${\rm Br}(\tau^{} \rightarrow \mu^{} \eta)_{\rm exp}$ is the
experimental upper limit on the branching ratio of $\tau^{} \rightarrow
\mu^{} \eta$.
It can be easily seen that the bound 
$(|\kappa_{32}^{\text{max}}|^{2})_{\tau\rightarrow \mu\eta}$ is rapidly relaxed 
in the region with a smaller $\tan\beta$ and a larger $m_{A}^{}$. 
In a similar way to  Eq.~\eqref{eq:bound-from-tmeta},
the maximal allowed value
$\left( |\kappa_{32}^{\text{max}}|^{2} \right)_{\text{mode}}$ can be
calculated for each mode.
The combined upper limit $|\kappa^{\text{max}}_{32}|^{2}$ is then given by
\begin{align}
\left|\kappa^{\text{max}}_{32}\right|^{2}
 \equiv
 \min \!
 \left\{ \!
  \left(
  \left|
   \kappa^{\text{max}}_{32}
  \right|^{2}
  \right)_{\tau\rightarrow\mu\eta} \! \! ,
  \left(
  \left|
   \kappa^{\text{max}}_{32}
  \right|^{2}
  \right)_{\tau \rightarrow \mu \gamma} \! \! , 
 \cdots \!
 \right\} \! .
\end{align}
We have found that  
$\tau^{}\rightarrow \mu\eta$ and 
$\tau^{}\rightarrow \mu\gamma$
give the strongest upper limits on $|\kappa_{32}|^{2}$
in a wide range of the parameter space\cite{kot}. 

We here consider LFV Higgs boson decay;
i.e., $h \rightarrow \tau^\pm \mu^\mp$.
Branching ratios for $h$ decay is 
calculated to be
\begin{align}
\hspace{-2mm}
{\rm Br}(h\rightarrow \tau^{\mp} \mu^\pm) =&
 \frac{m_{\tau}^{2} \cos^{2} \left(\alpha - \beta \right)}
 {16\pi v^{2} \cos^{4}\beta}
 \frac{\left| \kappa_{32} \right|^{2} m_{h}^{} }
 {\Gamma(h\rightarrow\text{all})}, 
\label{eq:Brhtm}
\end{align}
where $\Gamma (h \rightarrow \text{all}) $  
is the total width for $h$. 
The LFV Higgs boson decay can give restriction 
for LFV couplings 
\begin{align}
\left| \kappa_{32}^{\rm max} \right|^2 &=
 {\rm Br}(h \to \tau \mu)_{\rm exp}
\times \frac{3 m_b^2}{m_\tau^2} 
\frac{\cos^2\beta \sin^2\alpha}{\cos^2(\alpha-\beta)},
\end{align}
with the same way as rare tau decays. 

\begin{figure}
\begin{center}
\setlength{\unitlength}{1cm}
\begin{picture}(8,3.5)
\put(0.5,0){\includegraphics[width=7.5cm]{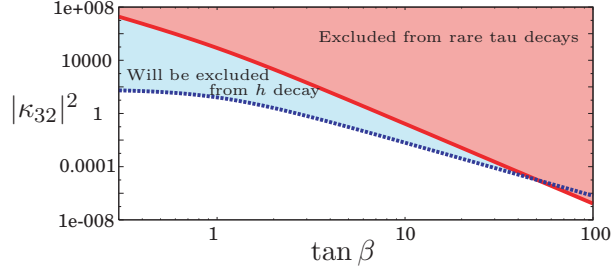}}
\put(0,1.7){$|\kappa_{32}|^2$}
\put(4,-0.2){$\tan\beta$}
\put(4.1,2.7){\tiny Excluded from rare tau decays}
\put(1.55,2.2){\tiny Will be excluded}
\put(2.65,2.0){\tiny from $h$ decay}
\end{picture}
\caption{The upper bounds on $|\kappa_{32}|^2$ from 
rare tau decays and also from LFV Higgs boson decay.}
\label{fig}
\end{center}
\end{figure}

In Fig.~\ref{fig}, we show the bound on $|\kappa_{32}|^2$ 
from rare tau decays and also from the Higgs boson decay 
$h \to \tau \mu$ as the function of $\tan \beta$. 
We take the upper limit 
${\rm Br}(h \to \tau \mu)_{\rm exp} = 0.8 \times 10^{-3}$, 
which is expected to be tested at the future linear collider~\cite{Osaka}.
The parameters are taken to be $m_{h}^{} =120$ GeV, 
$m_A^{}=m_{H}^{} = m_{H^{\pm}} = 350$ GeV and 
$\sin(\alpha -\beta) = -0.9999$. 
We have found that in the smaller $\tan\beta$ region 
LFV coupling $|\kappa_{32}|$ can be further constrained 
from LFV Higgs boson decay 
in comparison with that from rare tau decays~\cite{kot}.
\section{Conclusions}
\label{Sec:conclusion}
Lepton flavor violating decays of Higgs bosons 
have been studied in the framework of the LFV THDM.
The LFV coupling $|\kappa_{32}|^2$ are bounded 
from above by using the current data for rare tau LFV decays. 

It has been found that among the rare tau decay data 
those for 
$\tau^{}\rightarrow \mu\eta$ and 
$\tau^{}\rightarrow \mu\gamma$
give the most stringent upper limits on $|\kappa_{32}|^{2}$
in a wide range of the parameter space. 
In the large $\tan\beta$ region,  
the upper limit  on $|\kappa_{32}|^2$ due to the rare tau 
decay data turns out to be substantial. 
The upper limit would be improved in future by about one order of magnitude 
at the experiment at (super) B factories.
For smaller values of $\tan\beta$, the upper limit
is rapidly relaxed, and no more substantial constraint is obtained from 
the rare tau decay results. 

We have shown that a search for the LFV decays 
$h \rightarrow \tau^\pm \mu^\mp$ can be useful 
to further constrain the LFV Yukawa couplings
at future collider experiments. 
LFV decay of the lightest Higgs boson 
can be one of the important probes to find the evidence for the 
extended Higgs sector when the SM-like situation would 
be preferred by the data at forthcoming collider experiments.
The branching ratio for $h \to \tau^\pm \mu^\mp$ can be 
larger than $\mathcal{O}(10^{-3})$ 
except for the high $\tan\beta$ region. 
At future collider experiments, such a size of 
the branching fractions can be tested. 
Therefore, we conclude that the search of LFV in the Higgs 
boson decay at future colliders 
can further constrain the LFV Yukawa couplings 
especially in the relatively small $\tan\beta$ region, 
where rare tau decay data cannot reach.



\end{document}